\documentclass{ifacconf}
\usepackage{graphicx}
\usepackage{amssymb}
\usepackage{mathrsfs}
\usepackage{algorithm}
\usepackage{algpseudocode}
\usepackage{natbib}
\usepackage{bm}
\usepackage{nicefrac}
\usepackage{amsmath}
\usepackage{commath}
\usepackage[utf8]{inputenc}
\usepackage{enumerate}
\usepackage[implicit=false]{hyperref}
\usepackage{tikz}
\usepackage{tabularx}
\usepackage{arydshln}
\usetikzlibrary{arrows}
\usetikzlibrary{shapes.geometric, shapes.multipart}
\tikzstyle{block}=[rectangle,text centered]
\usepackage{mathtools}
\DeclareMathOperator*{\argmax}{argmax}

\newtheorem{thrm}{Theorem}
\newtheorem{remark}{Remark}
\newtheorem{lemma}{Lemma}
\newtheorem{assumpt}{Assumption}

\theoremstyle{definition}
\begin{document}

\begin{frontmatter}

\title{Generalised Regret Optimal Controller Synthesis for Constrained Systems} 
\author[First]{Alexandre Didier} 
\author[First]{Melanie N. Zeilinger} 
\address[First]{Institute for Dynamic Systems and Control (IDSC), ETH Zurich, 8092 Zurich, Switzerland (e-mail: \{adidier, mzeilinger\}@ethz.ch)}
\begin{abstract}                
This paper presents a synthesis method for the generalised dynamic regret problem, comparing the performance of a strictly causal controller to the optimal non-causal controller under a weighted disturbance. This framework encompasses both the dynamic regret problem, considering the difference of the incurred costs, as well as the competitive ratio, which considers their ratio, and which have both been proposed as inherently adaptive alternatives to classical control methods. Furthermore, we extend the synthesis to the case of pointwise-in-time bounds on the disturbance and show that the optimal solution is no worse than the bounded energy optimal solution and is lower bounded by a constant factor, which is only dependent on the disturbance weight. The proposed optimisation-based synthesis allows considering systems subject to state and input constraints. Finally, we provide a numerical example which compares the synthesised controller performance to $\mathcal{H}_2$- and $\mathcal{H}_\infty$-controllers.
\end{abstract}
\begin{keyword}
Predictive Control, Optimal control theory, Constrained control, Convex optimization, Robust controller synthesis
\end{keyword}
\end{frontmatter}
\section{INTRODUCTION} 
In recent years, regret minimisation has gained popularity as an adaptive controller performance metric. The regret of a system is defined as the difference of the incurred cost with respect to the optimal cost of a counter-factual system which has access to all future disturbances. This performance metric is adaptive in the sense that algorithms which are designed to minimise regret make use of the observed disturbances and potentially unknown, adversarial costs in order to adapt their behaviour. 

In policy regret, the controller which the designed algorithm is compared to is typically restricted to the class of linear feedback laws and convex costs are considered, see e.g. \cite{agarwal2019online}. The causal disturbance feedback policy, which is applied to the system, is updated at every time step using iterative update methods like gradient descent or Newton's method with the observed disturbance and cost in an online convex optimisation manner. The goal is then to show a sublinear regret with respect to the task horizon, as is done, e.g. in \cite{foster2020logarithmic} in the linear quadratic regulation case or in \cite{boffi2021regret} for nonlinear dynamics. As the regret is sublinear in the task horizon, the time-averaged cumulative regret vanishes with increasing task horizon, implying that the proposed algorithm learns the optimal benchmark controller in the restricted policy class. 

Another popular notion of regret is that of dynamic regret. As opposed to policy regret, the benchmark which is considered is not restricted to a policy class and can therefore be any control sequence. Dynamic regret minimisation in dynamic systems was first considered in \cite{goel2021regret} in the linear quadratic regulation case. A closed-form solution is derived for the dynamic regret optimal controller and it is shown that the incurred dynamic regret of the proposed controller is bounded by a linear factor in the disturbance energy. Note that compared to the algorithms typically presented in policy regret minimisation, this implies that the controller which minimises the cumulative dynamic regret does not necessarily incur sublinear regret in the considered task horizon. However, the optimality with respect to the specified control objective as well as the possibility to derive the control law in an offline manner enables the proposed method to be considered as an alternative to classical control methods such as $\mathcal{H}_2-$ and $\mathcal{H}_\infty$-control, as shown, e.g. in \cite{goel2021regret} and \cite{martin2022safe}. While these classical methods typically suffer from conservativeness when the disturbance does not match the underlying assumption, i.e. Gaussian and adversarial, respectively, the optimal dynamic regret controller is shown to outperform both the $\mathcal{H}_2-$ and the $\mathcal{H}_\infty$-controller for many possible disturbance signals due to its inherently adaptive nature. Similar results are achieved when minimising the competitive ratio, which considers the ratio of the costs instead of the difference, as shown in \cite{goel2021competitive}. The main limitation of such closed-form solutions lies in not being able to account for additional structure in the given control task, like additional bounds on the disturbance signal, state constraints which can be safety-critical or hard actuator limitations in the inputs. The aim of this paper is to provide a framework for synthesising controllers for the generalised dynamic regret problem for constrained systems.

\textit{Related work.} In \cite{li2021safe}, an algorithm minimising policy regret with respect to the optimal non-causal linear feedback controller, which fulfils state and input constraints for every disturbance realisation, is considered and in \cite{nonhoff2021online}, online convex optimisation is used to minimise dynamic regret for constrained systems with unknown cost functions and no disturbances. A dynamic regret bound of the $\mathcal{H}_\infty$-controller is provided in \cite{karapetyan2022regret}. A solution for the finite horizon dynamic regret optimal controller was derived in \cite{goel2021regret} and for the infinite horizon case in \cite{sabag2021regret}. A closed-form solution for the competitive ratio optimal controller was derived for finite horizon in \cite{goel2021competitive} and the infinite horizon case was considered in \cite{goel2021competitive} and \cite{sabag2022optimal}. Additionally, the generalised regret and its solution in an operator-theoretic form are given in \cite{sabag2022optimal} for the infinite horizon case. In \cite{martin2022safe} and \cite{didier2022system}, a semi-definite program (SDP) based synthesis framework is provided, which allows deriving the dynamic regret optimal controller and allows incorporating state and input constraints of the system. Furthermore, in \cite{didier2022system}, known initial conditions and pointwise-in-time bounds on the disturbance are considered, which allows synthesising controllers with no worse regret than for bounded energy and a suboptimality bound is provided on the approximate synthesis solution. 

\textit{Contribution.} In this paper, we propose a convex optimisation approach to the generalised dynamic regret problem. Compared to prior work in \cite{didier2022system}, the proposed approach encompasses not only the dynamic regret setting considered, e.g., in \cite{goel2021regret} and \cite{martin2022safe}, but also the competitive ratio setting as considered in in \cite{goel2021competitive}, as well as allowing to design a disturbance weight from problem specifications. We show how the optimal solution can be synthesised for bounded energy disturbances with adversarial or with known initial state. Furthermore, given ellipsoidal pointwise-in-time bounds on the disturbance, we propose a synthesis method which is no worse than the bounded energy solution. The potential suboptimality of the approach is lower bounded by a constant factor, which depends on the condition number of the general disturbance weighting matrix and recovers the suboptimality bound in the dynamic regret case proposed in \cite{didier2022system}. 
Finally, it is shown how state and input constraints can be incorporated into the synthesis.


\section{PROBLEM FORMULATION}\label{sec:PF} 
We consider discrete-time, linear time-varying system dynamics, with an unknown additive disturbance, of the form
\begin{equation}\label{eq:sysdyn}
x_{k+1}=A_k x_k+B_k u_k+E_k w_k,
\end{equation}
with states $x_k\in\mathbb{R}^n$, inputs $u_k\in\mathbb{R}^m$ and unknown disturbances $w_k\in\mathbb{R}^p$. For the considered task horizon $T$, the time-varying system matrices $A_k\in\mathbb{R}^{n\times n}$, $B_k\in\mathbb{R}^{n\times m}$ and $E_k \in\mathbb{R}^{n\times p}$ are known and it holds that the matrices $E_k$ have full row rank for all time steps.

We denote the state, input and disturbance sequences over the task horizon $T$ as
    $\mathbf{x} = [ x_0^\top \;  x_1^\top \; \cdots \; x_{T}^\top ]^\top$,
    $\mathbf{u} = [ u_0^\top \; u_1^\top \; \cdots \; u_{T}^\top]^\top$ and 
    $\mathbf{w} = [ w_0^\top \; w_1^\top \; \cdots \; w_{T-1}^\top ]^\top$, respectively, and assume that the initial condition as well as a disturbance energy bound is known.
    \begin{assumpt}\label{ass:BE}
        The initial state $x_0$ and a disturbance energy bound $\omega\geq 0$ is known such that
        $$ \mathbf{w}\in\ell_2^\omega:=\{\mathbf{w}\in\mathbb{R}^{pT}\mid \mathbf{w}^\top \mathbf{w}\leq \omega\}. $$
    \end{assumpt}
Note that in Section~\ref{sec:GROC}, the adversarial or $0$ initial condition case is also discussed, where the disturbance energy bound, albeit finite, is not required to be known.

The system described in \eqref{eq:sysdyn} incurs a cumulative quadratic cost, which is denoted as
\begin{equation}
    J(x_0, \mathbf{u}, \mathbf{w}) = \sum_{k=0}^{T} x_k^\top Q_k x_k + u_k^\top R_k u_k,
\end{equation}
with cost matrices $Q_k\in\mathbb{R}^{n \times n}$ and $R_k\in\mathbb{R}^{m \times m}$ and it holds that $Q_k\succeq 0$ and $R_k \succ 0$, i.e. $Q_k$ are symmetric positive semi-definite and $R_k$ are symmetric positive definite. The cumulative cost can then be rewritten in terms of the state and input sequences as 
\begin{equation}\label{eq:costfunct}
        J(x_0, \mathbf{u}, \mathbf{w}) = \mathbf{x}^\top \mathcal{Q} \mathbf{x} + \mathbf{u}^\top \mathcal{R} \mathbf{u},
\end{equation}
where we introduce cost matrices $\mathcal{Q}{=}\textup{blkdiag}(Q_0, \dots, Q_T)$ and $\mathcal{R}{=}\textup{blkdiag}(R_0, \dots, R_T)$, with $\textup{blkdiag}(.)$ denoting the block diagonal matrix of the given matrices.

This paper addresses the problem of minimising the generalised dynamic regret, which is defined as 
\begin{equation}\label{eq:GenReg1}
   J(\bm{\delta},\mathbf{u}) - \min_{\tilde{\mathbf{u}}(\bm{\delta})} J(\bm{\delta}, \tilde{\mathbf{u}}(\bm{\delta})) \leq \mu  \bm{\delta}^\top \mathcal{W}_w \bm{\delta}\;\; \forall \mathbf{w}\in\ell_2^\omega,
\end{equation}
for the disturbance weight $\mathcal{W}_w \succ 0$, using $\bm{\delta}:=[x_0^\top \; \mathbf{w}^\top]^\top$, i.e. the aim is to find the optimal causal control sequence $\mathbf{u}^*$ and the corresponding, minimal generalised dynamic regret level $\mu^*$ for which \eqref{eq:GenReg1} holds. 
The problem therefore consists of minimising the difference between the incurred cumulative cost \eqref{eq:costfunct} of the system and the optimal control cost achieved by the non-causal control sequence which has access to all future disturbances $\tilde{\mathbf{u}}^*(\bm{\delta})$. This generalised framework is also considered in \cite{sabag2022optimal} in the infinite horizon case, where an operator-theoretic solution is provided. Note that in the finite horizon case which is considered in this paper, the weights on states and inputs can be directly incorporated in the time-varying $Q_k$ and $R_k$ matrices such that only the disturbance weight $\mathcal{W}_w$ is considered in \eqref{eq:GenReg1}. 

As shown in \cite{goel2020power}, the optimal non-causal control sequence is given by
\begin{equation}\label{eq:NCCont}
     \tilde{\mathbf{u}}^*=-(\mathcal{R}+F^\top \mathcal{Q} F)^{-1}F^\top\mathcal{Q}G \bm{\delta},
\end{equation}
where $F\in\mathbb{R}^{n(T+1){\times} m(T+1)}$ and $G\in\mathbb{R}^{n(T+1){\times}(n+pT)}$ are the causal operators describing the state sequence
\begin{equation}
    \mathbf{x}=F\mathbf{u} + G\bm{\delta}.
\end{equation}
The optimal non-causal cost which is incurred by the control sequence \eqref{eq:NCCont} is given by 
\begin{equation}\label{eq:NCOpt}
    J^*(\bm{\delta}) = \bm{\delta}^\top G^\top \mathcal{Q} (\mathcal{I}+F\mathcal{R}^{-1}F^\top \mathcal{Q})^{-1}G\bm{\delta}=:\bm{\delta}^\top \mathcal{O} \bm{\delta},
\end{equation}
where $\mathcal{I}$ denotes the identity matrix of appropriate dimension. This closed-form solution for the optimal non-causal cost allows for a reformulation of \eqref{eq:GenReg1} for the optimal generalised dynamic regret level $\mu^*$ as
\begin{equation}\label{eq:GenReg2}
    J(\bm{\delta},\mathbf{u}^*) - \bm{\delta}^\top \mathcal{O}\bm{\delta}\leq \mu^* \bm{\delta}^\top \mathcal{W}_w\bm{\delta} \;\; \forall \mathbf{w}\in\ell_2^\omega.
\end{equation}

We note that this generalised framework encompasses both the dynamic regret problem as considered in \cite{goel2021regret}, \cite{didier2022system} and \cite{martin2022safe} given adversarial noise through the choice $\mathcal{W}_w=\mathcal{I}$, i.e. the worst-case regret is given by
\begin{equation}
    \textup{Regret}(x_0, \mathbf{u}) = \max_{\Vert\mathbf{w}\Vert^2\leq\omega} J(\bm{\delta}, \mathbf{u}) - J^*(\bm{\delta}),
\end{equation}
as well as the competitive ratio problem considered in \cite{goel2021competitive} through the choice $\mathcal{W}_w = \mathcal{O}$, where the optimal competitive ratio given adversarial disturbances is given by
\begin{equation}
    \textup{CR}(x_0, \mathbf{u}) := \max_{\mathbf{w}\neq 0} \frac{J(\bm{\delta}, \mathbf{u})}{J^*(\bm{\delta})} = 1+\mu^*,
\end{equation}
where $\mu^*$ denotes the optimal performance level in \eqref{eq:GenReg2}.

In Section~\ref{sec:RegOptSLS}, we show how the generalised dynamic regret problem can be solved for bounded energy disturbance signals with known, as well as adversarial, initial conditions through an SDP and in Section~\ref{sec:StrucGROC}, we consider how additional pointwise-in-time bounds on the disturbances as well as state and input constraints can be incorporated into the proposed optimisation-based synthesis framework.

\section{OPTIMAL GENERALISED REGRET CONTROL VIA SYSTEM LEVEL PARAMETRISATION}\label{sec:RegOptSLS}
In this section, we first introduce the system level parametrisation (SLP), which reformulates the state and input sequences linearly with respect to the disturbance sequence. Through an affine constraint on the optimisation variables, it can be ensured that the state and input sequences are valid for the system dynamics \eqref{eq:sysdyn}. We then show how this parametrisation can be used to synthesise the generalised dynamic regret optimal controller through a convex optimisation problem.
\subsection{System Level Parametrisation}\label{sec:SLP}
The SLP allows synthesising $\mathcal{H}_2$- and $\mathcal{H}_\infty$-optimal controllers, as shown e.g. in \cite{anderson2019system}, as well as the dynamic regret optimal controller, as shown in \cite{martin2022safe} and \cite{didier2022system}, through convex optimisation problems and we show how it can be used in the generalised regret framework in Section~\ref{sec:GROC}. We first define
\begin{equation*}
    \mathcal{K} = \begin{bmatrix} K^{0,0}& 0 & \dots & 0 \\ K^{1,0} & K^{1,1} & \ddots & \vdots \\ \vdots & \ddots & \ddots & 0 \\ K^{T,0} & \dots & K^{T, T-1} & K^{T,T} \end{bmatrix},
\end{equation*}
as the causal, time-varying controller matrix, with $\mathbf{u}=~\mathcal{K}\mathbf{x}$, and the block diagonal system matrices as $\mathcal{A}=\textup{blkdiag}(A_0,$ $ {\dots},$$ A_T)$, $\mathcal{B}{=}\textup{blkdiag}(B_0,{\dots},B_T)$ and $\mathcal{E}{=}\textup{blkdiag}(\mathcal{I}, E_0, {\dots},$ $E_{T-1})$. The SLP is then given by
\begin{equation}\label{eq:SLP}
\begin{bmatrix}
    \mathbf{x} \\ \mathbf{u}
\end{bmatrix}\! {=}\! \begin{bmatrix}
    (\mathcal{I}{-}\mathcal{Z}(\mathcal{A}{+}\mathcal{B}\mathcal{K}))^{\!{-}1}\mathcal{E} \\ \mathcal{K}(\mathcal{I}{-}\mathcal{Z}(\mathcal{A}{+}\mathcal{B}\mathcal{K}))^{\!{-}\!1}\mathcal{E}
\end{bmatrix}\bm{\delta} {:=} \begin{bmatrix} \Phi_x \\ \Phi_u \end{bmatrix}\bm{\delta} {:=} \Phi \bm{\delta} {:=} \begin{bmatrix} \Phi_0 & \Phi_w \end{bmatrix}\bm{\delta},
\end{equation}
where $\mathcal{Z}$ is the downshift operator. For the system response $\Phi$, two different block row and block column decompositions into $\Phi_x$ and $\Phi_u$ and $\Phi_0$ and $\Phi_w$, respectively, are utilised when convenient. It then holds that the controller matrix $\mathcal{K}$ can be synthesised, given a system response $\Phi$, as 
\begin{equation} \label{eq:SLPCont}
    \mathcal{K} = \Phi_u \Phi_x^{-1}.
\end{equation}
Note that for a strictly causal controller matrix $\mathcal{K}$, the state and input system responses $\Phi_x$ and $\Phi_u$ are causal operators. Furthermore, it was shown in \cite[Theorem~2.1]{anderson2019system} and in \cite[Theorem~1]{chen2021system}, that for all possible system responses, the affine constraint
\begin{equation}\label{eq:SLPConstr}
    \begin{bmatrix} \mathcal{I}-\mathcal{Z}\mathcal{A} & -\mathcal{Z}\mathcal{B} \end{bmatrix} \Phi = \mathcal{E}  
\end{equation}
holds. The SLP therefore allows optimising over the system response $\Phi$, given that \eqref{eq:SLPConstr} holds and the controller which is applied to the system can be recovered using \eqref{eq:SLPCont}.

\subsection{Optimal Generalised Regret Controller}\label{sec:GROC}
In this section, we show how the optimal solution to the generalised regret problem \eqref{eq:GenReg1}
can be recovered from the solution of a convex semi-definite program (SDP). Given the SLP \eqref{eq:SLP}, the cumulative incurred cost of the system can be rewritten as a quadratic function of the initial state and the disturbances, leading to the generalised regret problem formulation with $\mathcal{C}:=\textup{blkdiag}(\mathcal{Q},\mathcal{R})$
\begin{equation}\label{eq:GenRegSLP}
\begin{split}
    &\bm{\delta}^\top (\Phi^\top \mathcal{C} \Phi - \mathcal{O}) \bm{\delta} \leq \mu \bm{\delta}^\top \mathcal{W}_w\bm{\delta} \;\; \forall \mathbf{w}\in\ell_2^\omega \\
    \Leftrightarrow&\bm{\delta}^\top (\Phi^\top \mathcal{C}\Phi - \mathcal{O} - \mu \mathcal{W}_w)\bm{\delta} \leq 0 \;\; \forall \mathbf{w}\in\ell_2^\omega.
    \end{split}
\end{equation}
The problem of finding the optimal generalised performance level $\mu^*$ and the corresponding control matrix $\mathcal{K}^*$ can therefore be reformulated as the following optimisation problem
\begin{subequations}
    \begin{align}
        \min_{\mu, \Phi} & \; \mu \label{eq:GenRegSLP1} \\
        \textup{s.t. } & \max_{\Vert\mathbf{w}\Vert^2\leq \omega} \bm{\delta}^\top (\Phi^\top \mathcal{C} \Phi - \mathcal{O} - \mu \mathcal{W}_w)\bm{\delta} \leq 0 \label{eq:GenRegSLP2} \\
        &\begin{bmatrix} \mathcal{I}-\mathcal{Z}\mathcal{A} & -\mathcal{Z}\mathcal{B} \end{bmatrix} \Phi = \mathcal{E}, \label{eq:GenRegSLP3}
    \end{align}
\end{subequations}
where we minimise the performance level $\mu$ in \eqref{eq:GenRegSLP1} such that it is valid, which is ensured through \eqref{eq:GenRegSLP2}, and the corresponding system response is valid for system \eqref{eq:sysdyn} through \eqref{eq:GenRegSLP3}. By reformulating the constraint \eqref{eq:GenRegSLP2} as a convex constraint with respect to the optimisation variables $\mu$ and $\Phi$, we can formulate the generalised dynamic regret problem as a convex optimisation problem. Before stating the main result, we note that the maximisation in constraint \eqref{eq:GenRegSLP2} can be written as a quadratically constrained quadratic program (QCQP) for a given disturbance energy bound $\omega$ as follows
\begin{equation} \label{eq:QCQP}
\begin{split}
    0\geq\max_{\mathbf{w}}\; &\mathbf{w}^\top (\Phi_w^\top \mathcal{C}\Phi_w - \mathcal{O}_{3} -\mu \mathcal{W}_{3}) \mathbf{w}  \\
    + &2\mathbf{w}^\top( \Phi_w^\top \mathcal{C}\Phi_0 - \mathcal{O}_{2} -\mu \mathcal{W}_{2}) x_0 \\ +&x_0^\top(\Phi_0^\top \mathcal{C} \Phi_0 - \mathcal{O}_{1} -\mu \mathcal{W}_{1}) x_0 \\
    \textup{s.t. } & \mathbf{w}^\top \mathbf{w} \leq \omega,
\end{split}
\end{equation}
where we define
\begin{equation*}
    \mathcal{O}=\begin{bmatrix}
        \mathcal{O}_{1} & \mathcal{O}_{2}^\top \\ \mathcal{O}_{2} & \mathcal{O}_{3}
    \end{bmatrix}
    \textup{ and }
    \mathcal{W}_w=\begin{bmatrix}
        \mathcal{W}_{1} & \mathcal{W}_{2}^\top \\ \mathcal{W}_{2} & \mathcal{W}_{3}
    \end{bmatrix}.
\end{equation*}

\begin{thrm}\label{thrm:GenReg} 
    The optimal generalised dynamic regret performance level under Assumption~\ref{ass:BE} is given by the optimal value of the SDP
    \begin{subequations} \label{eq:GenRegOpt}
    \begin{align}
            \min_{\mu, \lambda, \Phi} \; & \mu  \\
             \textup{s.t. } &\lambda \geq 0 \\ 
             &\begin{bmatrix} \mathcal{I}-\mathcal{Z}\mathcal{A} & -\mathcal{Z}\mathcal{B} \end{bmatrix} \Phi = \mathcal{E}\\
              & \hspace{-1.05cm}  \begin{bmatrix} x_0^\top (\mathcal{O}_{1} {+}\mu \mathcal{W}_{1})
 x_0 {-}\lambda \omega & x_0^\top (\mathcal{O}_{2}^\top{+}\mu \mathcal{W}_{2}^\top) & x_0^\top \Phi_0^\top\mathcal{C}^{\frac{1}{2}} \\ 
 (\mathcal{O}_{2}+\mu \mathcal{W}_{2})x_0 & \lambda \mathcal{I} {+} \mathcal{O}_{3} {+} \mu\mathcal{W}_{3} & \Phi_w^\top \mathcal{C}^{\frac{1}{2}} \\
 \mathcal{C}^\frac{1}{2} \Phi_0 x_0 &  \mathcal{C}^{\frac{1}{2}}\Phi_w & \mathcal{I}
            \end{bmatrix}{\succeq} 0 \label{eq:GenRegOptLMI}
            \end{align}
    \end{subequations}
    and the corresponding generalised dynamic regret optimal controller is given by 
    \begin{equation*}
        \mathcal{K}^* = \Phi_u^*\Phi_x^{* -1}.
    \end{equation*}
\end{thrm}
\begin{pf}
As Slater's condition is fulfilled by the QCQP \eqref{eq:QCQP}, we can write down the dual SDP, for which it is shown, e.g., in \cite{boyd2004convex}, that strong duality holds, i.e. the optimal value of the primal is equal to the optimal dual value. The constraint \eqref{eq:QCQP} is then equivalent to the constraints $\lambda\geq0$ and
\begin{equation*}
    \begin{split}
        &\begin{bmatrix}
            x_0^\top(\mathcal{O}_{1}+\mu \mathcal{W}_{1})x_0 - \lambda \omega & x_0^\top (\mathcal{O}_{2}^\top +\mu \mathcal{W}_{2}^\top) \\ (\mathcal{O}_{2} +\mu \mathcal{W}_{2})x_0 & \lambda \mathcal{I}+\mathcal{O}_{3}+\mu \mathcal{W}_{3}
        \end{bmatrix} \\
       - &\begin{bmatrix} x_0^\top \Phi_0^\top \\ \Phi_w^\top 
        \end{bmatrix} \mathcal{C}^{\frac{1}{2}} \mathcal{I} \mathcal{C}^{\frac{1}{2}} \begin{bmatrix} \Phi_0x_0 & \Phi_w
        \end{bmatrix} \succeq 0.
    \end{split}
\end{equation*}
Finally, by applying the Schur complement, see, e.g., \cite{vanantwerp2000tutorial}, the linear matrix inequality in \eqref{eq:GenRegOpt} is obtained and the generalised regret optimal controller can be computed through \eqref{eq:SLPCont}.\qed
\end{pf}
We note that the generalised regret optimal controller can be computed in polynomial time similarly to the dynamic regret setting, as the matrix inequality \eqref{eq:GenRegOptLMI} is linear in the optimisation variables and SDPs can be solved in polynomial time through interior point methods, see e.g. \cite{nesterov1994interior}. Note that the SDP for computing the dynamic regret optimal problem in \cite{didier2022system} is recovered from \eqref{eq:GenRegOpt} by redefining $\hat{\lambda}=\mu+\lambda$ and $\hat{\mu} = \mu (x_0^\top x_0 + \omega)$ as only the special case $\mathcal{W}_w=\mathcal{I}$ is considered.

In the special cases where the initial state $x_0=0$ or is adversarial, a simpler formulation to \eqref{eq:GenRegOpt} can be derived, for which knowledge of an explicit disturbance energy bound $\omega$ in Assumption~\ref{ass:BE} is no longer required. In fact, the generalised dynamic regret problem can then be reformulated as a generalised maximum singular value problem, similarly to the maximum singular value problem proposed in the dynamic regret case in \cite{martin2022safe} and \cite{didier2022system}.
We can rewrite the problem for $0$ initial condition as
        \begin{align*}
            & \mathbf{w}(\Phi_w^\top \mathcal{C}\Phi_w - \mathcal{O}_{3})\mathbf{w} \leq \mu \mathbf{w}^\top \mathcal{W}_{3} \mathbf{w} \; \forall \mathbf{w}  \\
            \Leftrightarrow & \tilde{\mathbf{w}} ^\top \mathcal{W}_{3}^{-\frac{1}{2}}(\Phi_w^\top \mathcal{C} \Phi_w - \mathcal{O}_{3} )\mathcal{W}_{3}^{-\frac{1}{2}}\tilde{\mathbf{w}} \leq \mu \tilde{\mathbf{w}}^\top \tilde{\mathbf{w}} \; \forall \tilde{\mathbf{w}},
        \end{align*}
    with $\tilde{\mathbf{w}}=\mathcal{W}_{3}^{\frac{1}{2}} \mathbf{w}$, from which it holds that 
    \begin{equation}\label{eq:2normGDR}
        \mu=\norm{\mathcal{W}_{3}^{-\frac{1}{2}}(\Phi_w^\top \mathcal{C} \Phi_w - \mathcal{O}_{3} )\mathcal{W}_{3}^{-\frac{1}{2}}}_{2\rightarrow 2}
    \end{equation}
    is a tight generalised regret performance level. 
    The minimisation of the generalised dynamic regret performance level is therefore equivalent to minimising the maximum singular value in \eqref{eq:2normGDR} and the corresponding synthesis problem is given by
    \begin{equation} \label{eq:GenRegNoInit}
        \begin{split}
            \min_{\mu, \Phi} \; & \mu \\ 
            \textup{s.t. } & \begin{bmatrix} \mu \mathcal{I} + \mathcal{W}_{3}^{-\frac{1}{2}}\mathcal{O}_{3}\mathcal{W}_{3}^{-\frac{1}{2}} & \mathcal{W}_{3}^{-\frac{1}{2}}\Phi_w^\top \mathcal{C}^{\frac{1}{2}} \\  
            \mathcal{C}^{\frac{1}{2}}\Phi_w\mathcal{W}_{3}^{-\frac{1}{2}} & \mathcal{I}
            \end{bmatrix} \succeq 0 \\
             &\begin{bmatrix} \mathcal{I}-\mathcal{Z}\mathcal{A} & -\mathcal{Z}\mathcal{B} \end{bmatrix} \Phi = \mathcal{E}.
        \end{split}
    \end{equation}
This reduced formulation follows directly from the matrix inequality formulation of the eigenvalue problem, i.e. $\mathcal{W}_{3}^{-\frac{1}{2}}(\Phi_w^\top \mathcal{C} \Phi_w - \mathcal{O}_{3} )\mathcal{W}_{3}^{-\frac{1}{2}}\preceq \mu \mathcal{I}$, or can be obtained from \eqref{eq:GenRegOpt} with $x_0=0$ and $\lambda = 0$. The adversarial initial condition case can be solved in an equivalent manner by considering the generalised dynamic regret problem
$\bm{\delta}^\top(\Phi^\top \mathcal{C}\Phi - \mathcal{O})\bm{\delta} \leq \mu \bm{\delta}^\top \mathcal{W}_{w} \bm{\delta} \; \forall\bm{\delta}$.
The corresponding synthesis problem is then equivalent to \eqref{eq:GenRegNoInit}, replacing the matrices $\mathcal{W}_{3}$, $\mathcal{O}_{3}$ and $\Phi_w$ with the full cost, weight and system response matrices $\mathcal{W}_w$, $\mathcal{O}$ and $\Phi$, respectively, and adjusting the dimensions of the identity matrices.

Finally, we note that any two controllers which are regret optimal for their respective disturbance weights achieve the same regret \textit{rates}, but with potentially different constant factors. To see this, we can consider the optimal system responses, disturbance weights and performance weights, $\Phi^*_1$, $\mathcal{W}_{w,1}$ and $\mu^*_1$ and $\Phi^*_2$, $\mathcal{W}_{w,2}$ and $\mu^*_2$, respectively. It then holds that 
\begin{align*}
    &\bm{\delta}^\top(\Phi^{*\top}_1\mathcal{C}\Phi^*_1{-}\mathcal{O})\bm{\delta}\leq \mu_1^* \bm{\delta}^\top\mathcal{W}_{w,1}\bm{\delta}\quad \forall\mathbf{w}\in\ell_2^\omega  \\
    \Leftrightarrow &\bm{\delta}^\top(\Phi^{*\top}_1\mathcal{C}\Phi^*_1{-}\mathcal{O})\bm{\delta}\leq \mu_1^* \sigma_{\textup{max}}(\mathcal{W}_{w,1}) \bm{\delta}^\top \bm{\delta}\quad \forall\mathbf{w}\in\ell_2^\omega  \\
    \Leftrightarrow &\bm{\delta}^\top(\Phi^{*\top}_1\mathcal{C}\Phi^*_1{-}\mathcal{O})\bm{\delta}\leq \mu_1^* \frac{\sigma_{\textup{max}}(\mathcal{W}_{w,1})}{\sigma_{\textup{min}}(\mathcal{W}_{w,2})}  \bm{\delta}^\top\mathcal{W}_{w,2}\bm{\delta} \quad\forall\mathbf{w}\in\ell_2^\omega 
\end{align*}

This implies that the dynamic regret optimal controller, i.e. using $\mathcal{W}_{w,1}=\mathcal{I}$, and the competitive ratio optimal controller, i.e. using $\mathcal{W}_{w,2}=\mathcal{O}$, achieve the same rate, but with different constant factors as we have, by optimality, $\mu^*_2\leq \mu^*_1\frac{1}{\sigma_{\textup{min}}(\mathcal{O})}$, as discussed in, e.g., \cite{andrew2013tale}.

\section{Structured and constrained generalised dynamic regret problems}\label{sec:StrucGROC}
In the previous section, we showed how the optimal solution to generalised regret problems can be synthesised through a convex semi-definite program for disturbances, that are $\ell_2$-signals and where the disturbance energy is assumed to be known if the initial state is used in the optimisation \eqref{eq:GenRegOpt}. 
While this disturbance formulation captures many relevant cases, more information about the disturbance is often assumed in practice in the form of a bound on the disturbance energy at every individual time step. When considering the worst case regret performance bound, such knowledge can significantly reduce the domain in the maximisation over the disturbance. In this section, we investigate how pointwise-in-time bounds on the disturbance can be exploited in the regret minimising synthesis problem. Furthermore, we show how state and input constraints can be directly incorporated into the proposed optimisation-based framework as shown in Section~\ref{sec:Constr}.

\subsection{Pointwise bounded disturbances}\label{sec:PWB}
We consider pointwise-in-time bounded disturbances, where the disturbance sets are assumed to be ellipsoids. Note that all the results presented in this and the next section can trivially be extended to time-varying bounds, but for ease of readability, we restrict the discussion to constant bounds.
\begin{assumpt}\label{ass:pwbell}
The disturbance $w_k$ in \eqref{eq:sysdyn} lies in a known, compact ellipsoid at every time step, i.e. 
$$w_k\in\mathbb{W}=\{w\in\mathbb{R}^{p} | \; w^\top P w\leq 1\},\ \forall k=0,\dots,T-1, $$
with $P \in\mathbb{R}^{p\times p}, P\succ0$.
\end{assumpt}
As the domain of the potentially adversarial disturbance is now assumed to be restricted to the $T$-times Cartesian product $\mathbb{W}^T=\mathbb{W}\times \dots \times \mathbb{W}$, the generalised dynamic regret problem which we would like to solve is given by
\begin{equation}\label{eq:GenRegPWB} 
J(\bm{\delta},\mathbf{u}) - J^*(\bm{\delta}) \leq \bar{\mu}\bm{\delta}^\top \mathcal{W}_w \bm{\delta} \quad \forall \mathbf{w}\in\mathbb{W}^T,
\end{equation}
and the optimal performance level which fulfils the inequality is denoted $\bar{\mu}^*_{\mathrm{PWB}}$. 
\begin{remark}
Note that for polytopic pointwise-in-time bounded disturbances $w_k\in\mathbb{W}$, the optimal solution can be computed by adding a constraint $\bm{\delta}_i^\top (\bar{\Phi}^\top \mathcal{C}\bar{\Phi}-\mathcal{O} -\bar{\mu}\mathcal{W}_w)\bm{\delta}_i\leq 0$ with $\bm{\delta}_i=[x_0^\top \mathbf{w}_i^\top]^\top$ for every vertex $\mathbf{w}_i$ of $\mathbb{W}^T$. However, this results in an exponential number of constraints in the time horizon, i.e. $n_w^T$, where $n_w$ is the number of vertices of $\mathbb{W}$.
\end{remark}
Similarly to the procedure described in Section~\ref{sec:GROC}, we can rewrite the inequality condition as a maximisation over all $\mathbf{w}\in\mathbb{W}^T$, which is equivalent to the quadratic constraints $\mathbf{w}^\top \mathcal{P}_i \mathbf{w}$ with the matrices $\mathcal{P}_i$ containing $P$ on the $i$-th block diagonal element and $0$ in all other entries. By considering the dual formulation to this problem, it is again possible to arrive at a semi-definite program which is convex in the optimisation variables. However, while strong duality holds for QCQPs with one single quadratic constraint, by introducing pointwise bounds at every time step, we arrive at $T$ individual quadratic constraints such that strong duality no longer holds in general, see, e.g., \cite{polik2007survey} for a counterexample with just two quadratic constraints. Nonetheless, it is possible to show that even if a suboptimal solution is synthesised, the generalised performance level obtained when considering pointwise-in-time bounded disturbances is no worse than the one obtained from considering an $\ell_2$-disturbance with equivalent disturbance energy, i.e. using the maximum disturbance energy $\omega = \frac{T}{\sigma_{\textup{min}}(P)}$ 
in the formulation \eqref{eq:GenRegOpt}.

\begin{prop}\label{prop:GenRegPWB}
    A suboptimal generalised dynamic regret performance level 
in \eqref{eq:GenRegPWB} under Assumption~\ref{ass:pwbell} is given by
    \begin{subequations}\label{eq:GenRegOptPWB}
    \begin{align}
     \bar{\mu}^* =   \min_{\bar{\mu}, \bar{\lambda}_i, \bar{\Phi}} & \; \bar{\mu} \\
             \textup{s.t. } &\bar{\lambda}_i \geq 0 \; \forall i =1,\dots, T \\ 
             &\begin{bmatrix} \mathcal{I}-\mathcal{Z}\mathcal{A} & -\mathcal{Z}\mathcal{B} \end{bmatrix} \bar{\Phi}= \mathcal{E}\\ 
  & \hspace{-1.7cm} \left[ \begin{matrix} x_0^{\!\top} (\mathcal{O}_{1} {+}\bar{\mu} \mathcal{W}_{1})
 x_0{-}\!\!\sum\limits_{i=1}^{T}\!\bar{\lambda}_i & x_0^\top (\mathcal{O}_{2}^\top{+}\bar{\mu} \mathcal{W}_{2}^\top) & x_0^\top \bar{\Phi}_0^\top\mathcal{C}^{\frac{1}{2}}\\ 
 (\mathcal{O}_{2}+\bar{\mu} \mathcal{W}_{2})x_0 & \!\sum\limits_{i=1}^{T}\!\bar{\lambda}_i\mathcal{P}_i {+} \mathcal{O}_{3} {+} \bar{\mu}\mathcal{W}_{3} & \bar{\Phi}_w^\top \mathcal{C}^{\frac{1}{2}} \\
 \mathcal{C}^\frac{1}{2} \bar{\Phi}_0 x_0 &  \mathcal{C}^{\frac{1}{2}}\bar{\Phi}_w & \mathcal{I}\end{matrix}\right]{\succeq}0, \nonumber \\
 \label{eq:GenRegOptPWBLMI}
            \end{align}
    \end{subequations}
    with the corresponding controller given by $\bar{\mathcal{K}}^*=\bar{\Phi}_u^*\bar{\Phi}_x^{* -1}$ and it holds that the true regret generalised performance level $\bar{\mu}^*_{\mathrm{PWB}}$ fulfils
    \begin{equation}\label{eq:GenRegUpper}\bar{\mu}^*_{\mathrm{PWB}}\leq \bar{\mu}^* \leq \mu^*,\end{equation}
    where $\mu^*$ denotes the optimal solution of \eqref{eq:GenRegOpt}.
\end{prop}
\begin{pf} The SDP formulation follows similarly to Theorem~\ref{thrm:GenReg} directly from the dual formulation of the maximisation of the QCQP with the $T$ constraints $\mathbf{w}^\top \mathcal{P}_i \mathbf{w}\leq 1$. By using Schur complement, the linear matrix inequality \eqref{eq:GenRegOptPWBLMI} is obtained and $\bar{\mu}_{\textup{PWB}}\leq\bar{\mu}^*$ follows from Lagrangian duality. In order to show the upper bound $\bar{\mu}^*\leq \mu^*$, we make use of the solution of \eqref{eq:GenRegOpt} with $\omega=\frac{T}{\sigma_{\mathrm{min}}(P)}$, denoted by $\lambda^*$ and $\Phi^*$, as a feasible solution in \eqref{eq:GenRegOptPWB}. By setting $\bar{\lambda}_1=\bar{\lambda}_2=\dots=\bar{\lambda}_T=\frac{\lambda^*\omega}{T}$, $\bar{\Phi}=\Phi^*$ and $\bar{\mu}=\mu^*$, it holds that \eqref{eq:GenRegOptPWB} is feasible as $\sum_{i=1}^{T}\bar{\lambda}_i\mathcal{P}_i = \frac{\omega\lambda^*}{T} \sum_{i=1}^{T}\mathcal{P}_i= \frac{\lambda^*}{\sigma_{\textup{min}}(P)}\sum_{i=1}^{T}\mathcal{P}_i \succeq \lambda^*\mathcal{I}$. As $\bar{\mu}$ is minimised the inequality with respect to the optimal solution of \eqref{eq:GenRegOpt} holds. \qed
\end{pf}
By solving \eqref{eq:GenRegOptPWB}, we can therefore synthesise a controller which performs at least as good as the generalised regret optimal controller for $\ell_2$-disturbances in \eqref{eq:GenRegOpt}. However, the optimal solution, which is recovered from the SDP does not necessarily optimally solve the generalised dynamic regret problem for pointwise bounded disturbances as only weak duality holds for the provided formulation. Due to weak duality, the obtained regret performance level is always an upper bound of the regret which is truly incurred. In order to provide bounds on the suboptimality of the recovered solution, we can make use of existing results of QCQP approximations in \cite{ye1999approximating}, which was reformulated in \cite{ben2001lectures}. 

\begin{lemma}\label{lem:DualApprox}
    Consider an optimisation problem of the form 
    \begin{equation*}
    \begin{split}
     p^* = \max_{z} &\; z^\top M_0 z \\
        \textup{s.t. } & z^\top M_i z\leq c_i  \; \forall i=0,\dots, q.
        \end{split}
    \end{equation*}
    Then it holds that if
    \begin{itemize}
        \item[1)] The matrices $M_1,\dots, M_q$ commute with each other,
        \item[2)] Slater's condition holds and there exists a combination of the matrices $M_1,\dots, M_q$ with non-negative coefficients which is positive definite,
        \item[3)] $M_0\succeq 0$,
    \end{itemize}
    the optimal value of
    its dual $q^*$
    fulfils the following suboptimality bound
    $$\frac{2}{\pi} d^* \leq p^*. $$
\end{lemma}
\begin{pf} The proof is given in \cite[Theorem 2]{ye1999approximating} and \cite[Proposition 4.10.5]{ben2001lectures}.\qed
\end{pf}
While this result is directly applicable for the dynamic regret case with $\mathcal{W}_{w}=\mathcal{I}$ in \cite{didier2022system}, the maximisation $\max_{\mathbf{w}\in\mathbb{W}^T} \bm{\delta}^\top (\Phi^\top \mathcal{C} \Phi - \mathcal{O} - \mu \mathcal{W}_w)\bm{\delta}$ is not directly related to the generalised regret level $\mu$ through optimal value of the problem. In order to make use of the suboptimality bound, we therefore consider the following auxiliary dynamic regret maximisation
\begin{equation}\label{eq:AuxGenReg}
p^* =\frac{1}{\bm{\delta}^{*\top}\bm{\delta}^*}\max_{\mathbf{w}\in\mathbb{W}^T}\; \bm{\delta}^\top (\bar{\Phi}_{\mathrm{PWB}}^{*\top} \mathcal{C}\bar{\Phi}_{\mathrm{PWB}}^*-\mathcal{O})\bm{\delta},
\end{equation}
where $\bar{\Phi}_{\mathrm{PWB}}^*$ is an optimal system response corresponding to the optimal generalised dynamic regret level $\bar{\mu}^*_{\mathrm{PWB}}$ and $\bm{\delta}^*$ is given by the initial state and an adversarial disturbance which maximises the dynamic regret $\mathbf{w}^*\in\argmax_{\mathbf{w}\in\mathbb{W}^T}\bm{\delta}^\top (\bar{\Phi}_{\mathrm{PWB}}^{*\top} \mathcal{C}\bar{\Phi}_{\mathrm{PWB}}^*-\mathcal{O})\bm{\delta}$.
\begin{prop}\label{prop:BoundGenRegOptPWB}
Consider the solution $\bar{\mu}^*$ of the optimisation problem \eqref{eq:GenRegOptPWB}. It holds that the optimal generalised regret level $\bar{\mu}^*_{\mathrm{PWB}}$ fulfils 
\begin{equation}\label{eq:lowerbound}
    \frac{2}{\pi\kappa(\mathcal{W}_w)}\bar{\mu}^*\leq \bar{\mu}^*_{\mathrm{PWB}},
\end{equation}
where $\kappa(\mathcal{W}_w)=\frac{\sigma_{\mathrm{max}}(\mathcal{W}_w)}{\sigma_{\mathrm{min}}(\mathcal{W}_w)}$ is the condition number of $\mathcal{W}_w$.
\end{prop}
\begin{pf}
It holds that the conditions in Lemma~\ref{lem:DualApprox} are fulfilled for \eqref{eq:AuxGenReg} as 1) $\mathcal{P}_i\mathcal{P}_j=\mathcal{P}_j\mathcal{P}_i=0$ for $i\neq j$, 2) Slater's condition holds and $\sum_{i=1}^T\mathcal{P}_i\succ 0$ as $P\succ 0$ and 3) the incurred generalised dynamic regret is non-negative for all disturbance realisations, i.e. $\Phi^\top\mathcal{C}\Phi-\mathcal{O}\succeq 0$ for all $\Phi$ which fulfill \eqref{eq:SLPConstr}. 
It therefore holds that $\nicefrac{2}{\pi}d^*\leq p^*\leq d^*$, where $d^*$ denotes the optimal value of the dual formulation of \eqref{eq:AuxGenReg}. Furthermore, it holds that $\bm{\delta}^{*\top} (\bar{\Phi}_{\mathrm{PWB}}^{*\top} \mathcal{C}\bar{\Phi}_{\mathrm{PWB}}^*-\mathcal{O})\bm{\delta}^*=p^*\Vert \bm{\delta}^{*}\Vert_2^2\leq \bar{\mu}^*_{\mathrm{PWB}}\bm{\delta}^*\mathcal{W}_w\bm{\delta}^*\leq \bar{\mu}^*_{\mathrm{PWB}} \sigma_{\max}(\mathcal{W}_w)\Vert \bm{\delta}^{*}\Vert_2^2$, from which it follows that $\frac{2}{\pi\sigma_{\mathrm{max}}(\mathcal{W}_w)}d^*\leq\bar{\mu}_{\mathrm{PWB}}^{*}$. By using the optimal solution $d^*$ and $\bar{\lambda}_{d,i}^*$ of the dual of \eqref{eq:AuxGenReg} as well as $\bar{\Phi}_{\mathrm{PWB}}^*$ in \eqref{eq:GenRegOptPWB}, it holds that $\mu=\frac{d^*}{\sigma_{\mathrm{min}}(\mathcal{W}_w)}$ is a feasible solution as $\frac{d^*}{\sigma_{\mathrm{min}}(\mathcal{W}_w)}\mathcal{W}_w\succeq \begin{bmatrix} d^* & 0 \\ 0 & 0\end{bmatrix}$, which concludes the proof due to the minimisation over $\mu$ in \eqref{eq:GenRegOptPWB}.\qed
\end{pf}
The computed solution therefore admits a constant lower bound, which depends only on the choice of the disturbance weight matrix $\mathcal{W}_w$, and which reduces to the bound provided in \cite{didier2022system} for $\mathcal{W}_w=\mathcal{I}$. We note that the bound in \eqref{eq:lowerbound} implies that when selecting a disturbance weight $\mathcal{W}_w$, the proposed controller synthesis in \eqref{eq:GenRegOptPWB}, using pointwise-in-time bounds on the disturbance, suffers in approximation quality when the matrix $\mathcal{W}_w$ has a high condition number. However, the worst-case regret incurred will still be lower compared to using a bounded disturbance energy as shown in Proposition~\ref{prop:GenRegPWB}.
By considering the results in this section we can then formulate the following relationships between the presented synthesis problems.
\begin{thrm}
Consider the generalised regret problem \eqref{eq:GenRegPWB} and its corresponding optimal performance level $\bar{\mu}_{\mathrm{PWB}}^*$ for disturbances which verify Assumption~\ref{ass:pwbell}. The following relationship holds
\begin{equation}\label{eq:GRIneq}
    \frac{2}{\pi\kappa(\mathcal{W}_w)}\bar{\mu}^*\leq \bar{\mu}^*_{\mathrm{PWB}} \leq \bar{\mu}^*\leq \mu^*,
\end{equation}
where $\mu^*$ and $ \bar{\mu}^*$ are the solutions of \eqref{eq:GenRegOpt} and \eqref{eq:GenRegOptPWB}, respectively.
\end{thrm}
\begin{pf}
    The relationship \eqref{eq:GRIneq} follows directly from Proposition~\ref{prop:GenRegPWB} and \ref{prop:BoundGenRegOptPWB}.
    \qed
\end{pf}
\begin{remark}
    The full block S-Procedure, as used, e.g., in \cite{scherer2001lpv}, can also be used to derive a synthesis problem similar to \eqref{eq:GenRegOptPWB}. This results in an increased number of optimisation variables, but it encompasses the S-Procedure as a special case, implying that the optimal solution of \eqref{eq:GenRegOptPWB} can be shown to be feasible in the resulting synthesis problem. Therefore a no worse generalised dynamic regret performance level is incurred. However, the resulting performance level showed no improvement compared to \eqref{eq:GenRegOptPWB} in the conducted numerical example.  
\end{remark}
\subsection{State and input constrained systems}\label{sec:Constr}
As physical systems are often subject to state and input constraints, we show how they can be considered in the controller synthesis for any disturbance $\mathbf{w}\in\mathbb{W}^T$. Given the constraint sets
\begin{equation}\label{eq:stateinputconstr}
\mathbb{X}=\{x{\in}\mathbb{R}^n\;|\; H_x x\leq \mathbf{1}\}, \;\; \mathbb{U}=\{u{\in}\mathbb{R}^m \;|\; H_u u\leq \mathbf{1} \}
\end{equation}
we can reformulate the constraints in terms of the SLP as $H{_z}\Phi\bm{\delta}\leq\mathbf{1}$, $\forall \mathbf{w}\in\mathbb{W}^{T}$ with $H_z = \textup{blkdiag}(\mathcal{I}\otimes H_x, \mathcal{I}\otimes H_u)$. As shown in \cite{didier2022system}, for ellipsoidal constraints satisfying Assumption~\ref{ass:pwbell}, the state and input constraints can directly be included in problem \eqref{eq:GenRegOptPWB} in the form of the following constraints
\begin{align} 
    &\max_{\mathbf{w} \in \mathbb{W}^{T}} [H_z]_{i} \Phi \bm{\delta} \leq 1, \; \forall i \nonumber \\
    \Leftarrow &[H_z]_{i}\Phi_0x_0+ {\sum_{j=1}^{T+1}}\norm{[H_{z}]_{i} [[\Phi_{w}]]_{j}  P^{-\frac{1}{2}}}\leq 1, \; \forall i\label{eq:constrdual}
\end{align}
which follows by exploiting the definition of the dual norm as shown, e.g. in \cite{goulart2006optimization} and where $[H_z]_i$ denotes the $i$-th row of $H_z$ and $[[\Phi_w]]_j$ denotes the $j$-th block column of appropriate size of $\Phi_w$. 
Note that due to the additional constraints on the system, the definition of the best achievable controller would also have to be adapted, as the optimal non-causal cost in \eqref{eq:NCOpt} may be achieved by violating the constraints. One possibility, which was proposed in \cite{martin2022safe}, is to consider the optimal non-causal controller, which fulfills the constraints for all possible disturbance sequences. Such a benchmark can be computed by minimising the cost $\Phi^\top\mathcal{C}\Phi$ using non-causal, i.e. full block, system responses $\Phi$ subject to the constraints \eqref{eq:constrdual}. The new benchmark cost $\tilde{\mathcal{O}}=\Phi^{*\top} \mathcal{C}\Phi^*$ can then be used to replace $\mathcal{O}$ in \eqref{eq:GenRegOptPWB}.

\section{NUMERICAL EXAMPLE}
We consider a 3-DoF rocket model, describing movement on a 2D plane, which is linearised around the hover position, inspired by \cite{spannagl2021design}. The states consist of the positions, velocities, angle and angular velocities of the rocket on the 2D plane, resulting in 6 states, 2 inputs, and dynamics of the form \eqref{eq:sysdyn}\footnote{The numerical values of the model parameters and constraints are accessible in the code, as linked in the Data Availability Statement.}.
The rocket is controlled over a horizon of $T=25$, with a non-zero initial condition, cost matrices $\mathcal{Q}{=}\mathcal{I}$, $\mathcal{R}{=}\mathcal{I}$ and an ellipsoidal disturbance bound $w_k\in\mathbb{W}=\{w \mid w^\top P w \leq 1\}.$
The system is subject to polytopic state and input constraints given by \eqref{eq:stateinputconstr}.
We synthesise the optimal controllers $\mathcal{S}\mathcal{H}_2$, $\mathcal{S}\mathcal{H}_\infty$, where $\mathcal{S}$ denotes guaranteed constraint satisfaction similar to \cite[Section 2.2]{anderson2019system}, using the constraint reformulation \eqref{eq:constrdual}. The dynamic regret $\mathcal{SDR}$ and competitive ratio $\mathcal{SCR}$ optimal controllers are synthesised using \eqref{eq:GenRegOpt}. As the disturbance is subject to pointwise-in-time bounds, we make use of the ellipsoidal bounds in the synthesis \eqref{eq:GenRegOptPWB}, resulting in the controllers $\mathcal{SDR}_{\mathbb{W}}$ and $\mathcal{SCR}_{\mathbb{W}}$. The considered synthesis problems are solved using YALMIP and MOSEK, see \cite{lofberg2004yalmip} and \cite{andersen2000mosek}, respectively.
The normalised cumulative costs for a number of disturbance realisations, constructed similarly to the numerical example in \cite{martin2022safe}, and satisfying $w_k\in\mathbb{W}$ are provided in Table~\ref{tbl:NE1}. The disturbance realisations are drawn from a truncated Gaussian distribution on $\mathbb{W}$, denoted $\overline{\mathcal{N}}_{\mathbb{W}}$, as well as a uniform distribution, denoted $\mathcal{U}_\mathbb{W}$ with domain $\mathbb{W}$, for which the cost is averaged over 100 disturbance realisations. 
We note that the dynamic regret and competitive ratio optimal controllers for pointwise bounded ellipsoidal disturbances consistently outperform the other controllers except in the case of a truncated Gaussian or uniform distribution, where the $\mathcal{SH}_2$-controller performs best. Furthermore, $\mathcal{SDR}_{\mathbb{W}}$ and $\mathcal{SCR}_{\mathbb{W}}$ show a considerably lower generalised regret performance level than their bounded energy counterparts, with a reduction of $47.1\%$ from $10.41$ to $5.50$ in the dynamic regret and $24.6\%$ from $0.61$ to $0.46$ in the competitive ratio case, respectively. 
\begin{table}
\setlength{\tabcolsep}{4.7pt}
\caption{Normalised incurred cumulative cost}
\label{tbl:NE1}
\begin{center}
\begin{tabular}{|c|c|c|c|c|c|c|}
\hline
$\mathbf{w}$ & $\mathcal{SH}_2$ & $\mathcal{SH}_\infty$ & $\mathcal{SDR}$ & $\mathcal{SCR}$ & $\mathcal{SDR}_{\mathbb{W}}$ & $\mathcal{SCR}_{\mathbb{W}}$ 
\\
\hline
$\overline{\mathcal{N}}_{\mathbb{W}}$ &  1  &  2.009  &  1.058  &  1.245 & 1.001  & 1.165  \\\hline
$\mathcal{U}_{\mathbb{W}}$ & 1  &  2.047  &  1.061  & 1.241  & 1.003  & 1.162  \\\hline
constant & 1.360  &  3.565  &  1.206  &  1.082 &  1.082  &  1 \\\hline
sinusoidal &  1.093  &  2.757  &  1.088  &  1.071  &  1.001  &  1 \\\hline
sawtooth &  1.018 &  1.730  & 1.055   &  1.296  &  1  &  1.201 \\\hline
step & 1.100  &  1.738 &  1.078  &  1.305  &  1  &  1.208 \\\hline
stair & 1.105  & 3.863  &  1.163  &  1.035  &  1.073  & 1  \\\hline
\end{tabular}
\vspace{0.3cm}
\end{center}
\end{table}

\section{CONCLUSION}
In this work, we presented an SDP-based approach for synthesising the generalised dynamic regret optimal controller for bounded energy disturbances. The proposed work extends previous results in \cite{didier2022system} to a more general class of problems, incorporating in particular also the competitive ratio setting in \cite{goel2021competitive}. Additional assumptions on the disturbance, such as pointwise ellipsoidal bounds can be used in the synthesis to reduce conservativeness through a QCQP duality approach. A constant lower bound based on the condition number of the disturbance weight matrix is provided and it is shown how state and input constraints satisfaction can be guaranteed for the synthesised controllers. 
\section*{Data availability statement}
The code and data in this study are available in the ETH Research Collection: \\ \href{https://doi.org/10.3929/ethz-b-000606880}{https://doi.org/10.3929/ethz-b-000606880}.

\bibliography{root}
\end{document}